# Charge doping to flat AgF$_2$ monolayers in a chemical capacitor setup

Daniel Jezierski[1], Adam Grzelak[1]*, XiaoQiang Liu[2], Shishir Kumar Pandey[2], Maria N. Gastiasoro,[3] José Lorenzana,[3] Ji Feng[2,4] and Wojciech Grochala[1]*

[1]Center of New Technologies, University of Warsaw, 02089 Warsaw, Poland
[2]International Center for Quantum Materials, School of Physics, Peking University, Beijing, 100871, China
[3]Institute for Complex Systems (ISC), Consiglio Nazionale delle Ricerche, Dipartimento di Fisica, Università di Roma "La Sapienza", 00185 Rome, Italy
[4]Collaborative Innovation Center of Quantum Matter, Beijing, 100871, China

*a.grzelak@cent.uw.edu.pl, w.grochala@cent.uw.edu.pl

ABSTRACT

Flat monolayers of silver(II) fluoride, which could be obtained by epitaxial deposition on an appropriate substrate, have been recently predicted to exhibit very strong antiferromagnetic superexchange and to have large potential for ambient pressure superconductivity if doped to an optimal level. It was shown that AgF$_2$ could become a magnetic glue-based superconductor with a critical superconducting temperature approaching 200 K at optimum doping. In the current work we calculate the optimum doping to correspond to 14% of holes per formula unit, i.e. quite similar to that for oxocuprates(II). Furthermore, using DFT calculations we show that flat [AgF$_2$] single layers can indeed be doped to a controlled extent using a recently proposed "chemical capacitor" setup. Hole doping associated with formation of Ag(III) proves to be difficult to achieve in the setup explored in this work as it falls at verge of charge stability of fluoride anions and does not affect the d($x^2$–$y^2$) manifold . However, in the case of electron doping, manipulation of different factors – such as number of dopant layers and the thickness of the separator – permits fine tuning of the doping level (and concomitantly $T_C$) all the way from underdoped to overdoped regime (in a similar manner as chemical doping for the Nd$_2$CuO$_4$ analogue).

INTRODUCTION

The intense quest for copper(II) oxide analogs, which could be doped to achieve unconventional high-$T_C$ superconductivity, has resulted during the last decades in several types of materials proposed to be researched.[1] Among those, AgF$_2$ seems to host powerful analogies with oxocuprates. Indeed, it shows strong 2D antiferromagnetic superexchange and substantial covalence of Ag–F bonding[2] which reflects in hopping parameters of a three-band tight-binding model very similar to those of superconducting cuprates. Furthermore, recent work has predicted that flat silver(II) fluoride monolayers stabilized by epitaxy on certain metal fluoride surfaces show an immense increase in two-dimensional antiferromagnetic (AFM) interactions with respect to bulk AgF$_2$.[3] Assuming a magnetic mechanism it has been estimated that the critical superconducting temperature ($T_c$) could reach up to 195 K in such materials.[3] In a recent theoretical study three of us (LXQ, SKP, JF) have determined possibility of the occurrence of superconductivity in bulk AgF$_2$ from the first principles.[4] The optimum doping level to the bulk material has been found to correspond to 5% hole doping, a result quite different from that for cuprates, where both a hole- or electron-doping at the *ca.* 16% level leads to superconductivity.[5,6] The same study also reported an unconventional singlet d-wave superconducting pairing in bulk AgF$_2$ which is found to strengthen with a decreasing interlayer coupling, highlighting the importance of quasi-2D nature of the crystal structure.[4]

In cuprates such as La$_2$CuO$_4$, doping is most often realized either by metal-metal substitution in the charge reservoir layers (La→Ba for hole doping, Nd→Ce for electron doping), or via introduction of O (or F) adatoms or vacancies. On the other hand, [AgF$_2$] layer is formally a 001-type precursor, and it lacks any charge reservoir layer. Thus, a question remains that how hole- or electron-doping to this material could be realized in practice. One key observation related to AgF$_2$ is that this compound is an immensely strong oxidizer. Translation of this feature from chemistry to physics implies that it is an electron-hungry $d^9$ system. Indeed, in its desire to fill the $d$ electron shell, AgF$_2$ exhibits very large electron affinity. Consequently, the value of work function for bulk AgF$_2$ for (010) surface is as large as 7.76 eV which appears to be extraordinary high surpassing even that of fluorinated diamond (7.24 eV).[7] The corresponding value for a single flat [AgF$_2$] sheet in vacuum is comparable to the latter value, 7.21 eV.[7] In any case, a [AgF$_2$] sheet is characterized by a very large chemical potential and is a powerful electron sink. In comparison, work function of CuO is 5.30 eV, *i.e.* some 2 eV smaller. These properties suggest that e-doping of AgF$_2$ is a much more feasible prospect than h-doping. In fact, AgF$_2$ tends to vigorously react with any non-fluoride materials; many oxides, chlorides and other salts yield spectacular electron-transfer reactions upon contact with AgF$_2$.[8] Density functional theory (DFT) calculations show that if AgF$_2$ is deposited on a purely oxide surface such as MgO, unfavorable charge transfer processes occur, which result in destabilization of AFM state and strong structural deformation of AgF$_2$ monolayer, leading to a material that is effectively overdoped.[3] A similar phenomenon has been computationally shown in fluoride-oxide superlattices, where AgF$_2$ sheet was in direct contact with oxide layers.[9,10] Remarkably, in none of chemical experiments conducted over the last 20 years has AgF$_2$ formed a partially e-doped system; instead, in all cases phase separation leading to the mixture of AgF and AgF$_2$ resulted from reactions; this suggests that AgF$_2$ shows very strong tendency for localization and clustering of additional charge carriers. This is also supported by quantum-mechanical calculations exploring the possibilities of chemical doping of bulk AgF$_2$, i.e. through modification of fluorine content.[11] In that study, both e- and h-doping leads to defect localization and points to tendencies towards phase separation.[11] However, these tendencies were found to be less pronounced in hypothetical flat AgF$_2$ structures with much broader electronic bands – both in bulk and as monolayers,[12] which raises hopes for achieving doped states in epitaxially stabilized AgF$_2$.

The following work is divided into three sections. In the first part, we determine the leading pairing stability and the optimum doping level using a DFT band structure for a flat layer and a weak coupling approach. Not surprisingly given the similarity with cuprates, the leading pairing instability is found to be d-wave. In the second part, in order to achieve electron-doping, we use electron-hungry characteristics of untamed AgF$_2$ for its own doom. We argue that the degree of electron transfer from oxide substrate to AgF$_2$ electron sink can be finely controlled through spatial separation of AgF$_2$ and oxide by a variable number of fluoride layers; this is a so called "chemical capacitor" setup, which three of us (AG, JL, WG) have recently proposed.[13] As our DFT calculations show, manipulation of the thickness of oxide layers and the thickness of the fluoride separator between them and the [AgF$_2$] surface sheet, permits fine adjustment of the doping level up all the way from underdoped, via the optimum doped and up to the overdoped regime, thus allowing for superconducting $T_C$ tuning. Finally, the third part of this work summarizes our findings related to the possibility of hole-doping to flat AgF$_2$ monolayer. We have explored three possible approaches for introducing holes into flat layer of AgF$_2$ and investigated their effects on its structure, as well as magnetic and electronic properties of the system. Unfortunately, all three strategies lead to depopulation of Ag d($z^2$) orbital, which is undesirable from the point of view of high-$T_c$ superconductivity in this material.

The band structure used for the superconducting computation assumes the system is in the paramagnetic state but close to a magnetic instability. Instead, to estimate the doping that one can induce with a chemical capacitor setup, we assumed a magnetic state. The rationale of these two different choices is that for the determination of doping, the important quantities are the capacitance of the AgF$_2$ layer (or the electronic compressibility) and the difference of the chemical potential that the two layers would have if the interlayer Coulomb interaction is neglected[13]. Both quantities are very sensitive to the presence of the charge gap at half–filling. Furthermore, we know that above some critical doping long-range order would vanish and only a finite magnetic correlation length would remain. Since the energy is determined by short range correlations, we argue that the magnetic DFT state represents the energy better than a paramagnetic state which has completely wrong magnetic and on-site correlation energy. On the other hand, to study superconductivity one is interested in quasiparticle excitations close to the Fermi level which are best described by the band structure corresponding to the paramagnetic DFT solution.

COMPUTATIONAL METHODS

All calculations in this work were performed within density functional theory (DFT) approach as implemented in VASP software using projector augmented wave potentials[14–18] and GGA-type Perdew-Burke-Ernzerhof functional adapted for solids (PBEsol).[19]

Estimate of the optimum doping level for superconductivity in weak-coupling

For the assessment of optimal doping levels to an isolated AgF$_2$ monolayer with first principles, we utilized the following procedure. A 15x15x1 Gamma-centered k-point mesh and energy cutoff of 520 eV for the plane waves is considered within our calculations. Lattice parameters of monolayer AgF$_2$ on tetragonal RbMgF$_3$ as substrate from Ref.[3] are used for flat AgF$_2$, which are $\boldsymbol{a} = \boldsymbol{b} = 4.0547$ Å. An interlayer separation of 30 Å is considered to prevent interlayer coupling. *Ab initio* band structure was projected onto a Wannier based tight binding model using WANNIER90-VASP[20] interface implemented within VASP to obtain the hopping parameters. The Hubbard model was then set up by adding an onsite $U$ term to this tight binding model with single $d$ orbital at each Ag site in the basis and is given by,

$$H = H^{\text{TB}} + H^{\text{hub}} = \sum_{ij\sigma} t_{ij}\, c^\dagger_{i\sigma} c_{j\sigma} + U \sum_i n_{i\uparrow} n_{i\downarrow} \qquad (1)$$

Here $c_{i\sigma}$ annihilates an electron in d(x$^2$-y$^2$) orbital on Ag in $i^{th}$ unit cell with spin $\sigma$ and $t_{ij}$ is the hopping matrix. First, we use random phase approximation (RPA)[21–25] to analyze the magnetic instability of the system. The RPA charge and spin susceptibilities are given by,

$$\begin{aligned} \chi^c(q) &= [1 + \chi^0(q)U]^{-1}\chi^0(q) \\ \chi^s(q) &= [1 - \chi^0(q)U]^{-1}\chi^0(q) \end{aligned} \qquad (2)$$

where $q = (\boldsymbol{q}, \omega)$. In the above expression the bare susceptibility $\chi^0(q)$ is given by,

$$\chi^0(\boldsymbol{q},\omega) = -\frac{1}{N}\sum_{\boldsymbol{k}} \frac{f(\varepsilon_{\boldsymbol{k}} - \varepsilon_F) - f(\varepsilon_{\boldsymbol{k}+\boldsymbol{q}} - \varepsilon_F)}{\omega + \varepsilon_{\boldsymbol{k}} - \varepsilon_{\boldsymbol{k}+\boldsymbol{q}} + i0^+} \qquad (3)$$

where N is the number of $k$-points and $\varepsilon_k$ is the corresponding eigenvalue at a particular $\boldsymbol{k}$-point. $f(E)$ is the Fermi distribution function and $\varepsilon_F$ is the Fermi energy.

Under fluctuation exchange approximation (FLEX),[26,27] the pairing vertex is given by,

$$\Gamma(q) = \gamma U \chi^s(q) U - \frac{1}{2} U \chi^c(q) U + U \tag{4}$$

with $\gamma = \frac{3}{2}$ for the singlet channel and $\gamma = -\frac{1}{2}$ for the triplet channel. The following linearized Eliashberg equation was then solved to obtain the order parameter and superconducting pairing symmetry,

$$\lambda \phi(k) = -\frac{T}{N} \sum_q \Gamma(q) \phi(k-q) G(k-q) G(q-k) \tag{5}$$

Here, $\lambda$ is the eigenvalue indicating the pairing strength and eigenvector $\phi(k)$ is the order parameter representing the superconducting pairing symmetry. $G(k)$ is the Matsubara Green's function.

In the weak coupling regime, the pairing vertex is approximated to be frequency-independent. Thus, we finally obtain the following equation,

$$\lambda \phi(\boldsymbol{k}) = -\frac{1}{N} \sum_{\boldsymbol{k}'} \Gamma^\eta(\boldsymbol{k}, \boldsymbol{k}') F(\boldsymbol{k}') \phi(\boldsymbol{k}') \tag{6}$$

where,

$$F(\boldsymbol{k}) = -\frac{f(\varepsilon_{\boldsymbol{k}} - \varepsilon_F) + f(\varepsilon_{-\boldsymbol{k}} - \varepsilon_F) - 1}{\varepsilon_{\boldsymbol{k}} + \varepsilon_{-\boldsymbol{k}} - 2\varepsilon_F} \tag{7}$$

arising from summing the product of Green's functions over Matsubara frequencies.

We use a mesh of 48x48x1 for sampling of Brillouin zone in our calculations related to superconductivity. The temperature was fixed at 30 meV. Pairing symmetries ($\phi(\boldsymbol{k})$) can be identified from the character table of the corresponding symmetry group of normal state Hamiltonian, which is $D_{4h}$ in our case.

Calculations of the e- and h-doped systems

For the study of electron- and hole-doping scenarios in the chemical capacitor setup, we performed calculations with the same settings as in our previous studies of surfaces/heterostructures with AgF$_2$ monolayers.[3,13,28] Magnetic interactions in AgF$_2$ monolayer were evaluated with antiferromagnetic coupling constant $J_{2D}$ ("2D" refers to two-dimensional coupling in the monolayer), which was calculated using collinear configurations and broken symmetry method as $J_{2D} = E_{AFM} - E_{FM}$ (where $E_{AFM}$ and $E_{FM}$ are energies per silver of antiferromagnetic and ferromagnetic solutions, respectively). By this convention, $J_{2D}$ is negative in AFM systems. Throughout the text, "larger/smaller $J_{2D}$" is taken to mean larger/smaller in magnitude, *i.e.* in its absolute value. VESTA software was used for visualization of structures.[29] Electronic density of states (eDOS) and band structure graphs were plotted with p4vasp software[30] and Python code. In eDOS plots, energy values are normalized so that $E_{Fermi} = 0$. In the studied systems, only *z* atomic coordinates (perpendicular to the substrate) were optimized, with the exception of F atoms within AgF$_2$ monolayer, where all three (x, y and z) coordinates were relaxed (although in some systems they were still constrained by symmetry).

<u>Determination of the doping level in DFT</u>

For electron doping, we have determined the number of carriers integrating the normalized DOS until the Fermi level. For hole doped cases, this approach is not possible because of the overlap of several bands close to the Fermi level. In those specific cases, we have computed the doping by fitting the bands crossing the Fermi level with a tight binding model and using Luttinger sum rule.

RESULTS AND DISCUSSION

<u>OPTIMAL DOPING LEVEL</u> The relevant hopping terms, extracted from our tight binding fit of $ab\ initio$ band structure shown in Fig. 1, are nearest neighbor hopping $t_1$=−0.4 eV(-0.444 eV), second nearest neighbours hopping $t_2$=0.025 eV(0.0284eV) and third nearest neighbours hopping $t_3$=−0.049 eV (−0.0357 eV), where in parentheses are the hopping strength in a CuO$_2$ plane for La$_2$CuO$_4$ taken from Ref. [31]. One can find out a striking similarity between AgF$_2$ and La$_2$CuO$_4$ in terms of these hopping interactions. Hubbard $U$ is set to a value of 0.9 eV, almost twice of $t_1$, similar to Ref. [32].

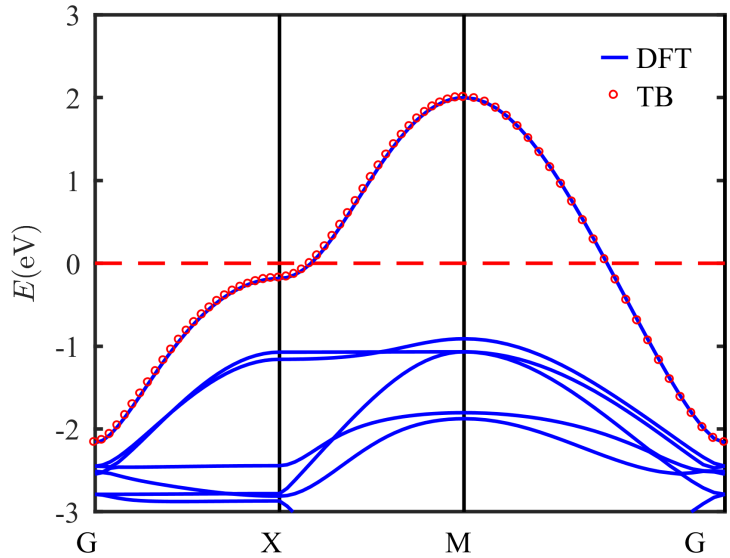

Figure 1. Non-spin polarized $ab\ initio$ band structure of flat AgF$_2$ projected onto a Wannier function based tight-binding model with Ag-$d(x^2-y^2)$ in the basis.

With this single band Hubbard model, within random phase approximation, spin instability appears when $\eta(\boldsymbol{q}) = \chi_0(\boldsymbol{q}, 0)U$ reach to unity at any $\boldsymbol{q}$ and manifest in terms of divergence of spin susceptibility [cf. Eq. (2)] causing a magnetic phase transition. The vector $\boldsymbol{q}$ and temperature $T$ at which $\eta(\boldsymbol{q})$ reaches to unity are called propagation vector and Néel temperature of the magnetic state.

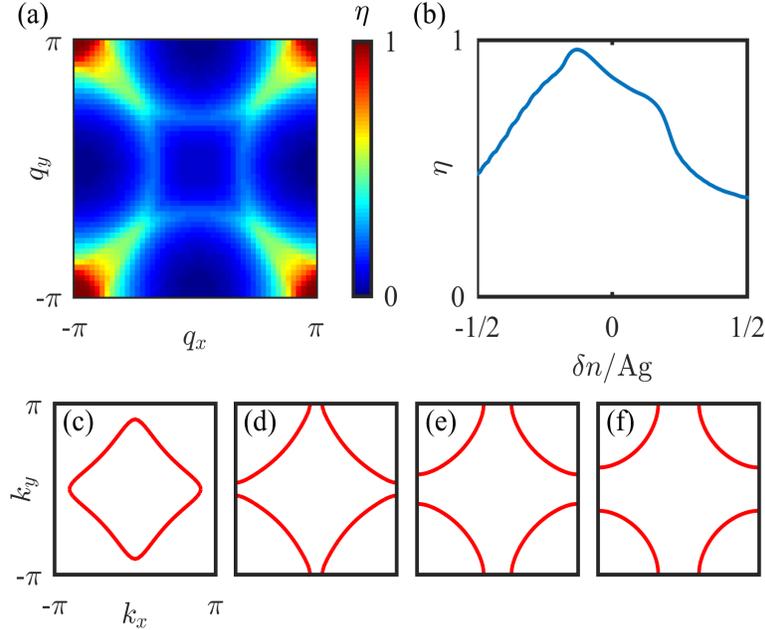

Figure 2. (a) $\eta(\boldsymbol{q})$ for undoped flat AgF$_2$. (b) The doping dependence of maximum value of $\eta(\boldsymbol{q})$. Fermi surface of doped flat AgF$_2$ with (c)$\delta n/$Ag $= -0.3$, (d)$\delta n/$Ag $= -0.15$, (e)$\delta n/$Ag $= 0$, (f)$\delta n/$Ag $= 0.15$.

Fig. 2(a) shows $\eta(\boldsymbol{q})$ in the 1st Brillouin zone for undoped flat AgF$_2$. Peak of $\eta(\boldsymbol{q})$ was found to be located at $(\pi,\pi)$, indicating a checkerboard-type antiferromagnetic ground state. The maximum value of $\eta(\boldsymbol{q})$ for various doping is shown in Fig. 2(b). We find the strongest spin fluctuation to occur at hole doping of $\delta n/$Ag $= -0.14$. This can be explained in terms of the Fermi surface nesting shown in Fig. 2(c-f). As one can observe that the Fermi surface is centered at $(\pi,\pi)$ for the undoped case (Fig. 2(e)). With electron doping, the hole pocket immediately starts to decrease (Fig. 2(f)), while for hole doping, the hole pocket first increases and reaches to largest nesting at ~ 14 % (Fig. 2(d)). With further hole doping, hole pocket is gradually changed to $\Gamma$ centered (Fig. 2(c)) electronic pocket with decrease in the pocket area when compared to 14% case. Having understood the weak coupling magnetic instability, we went on to examine the superconducting properties of flat AgF$_2$.[27]

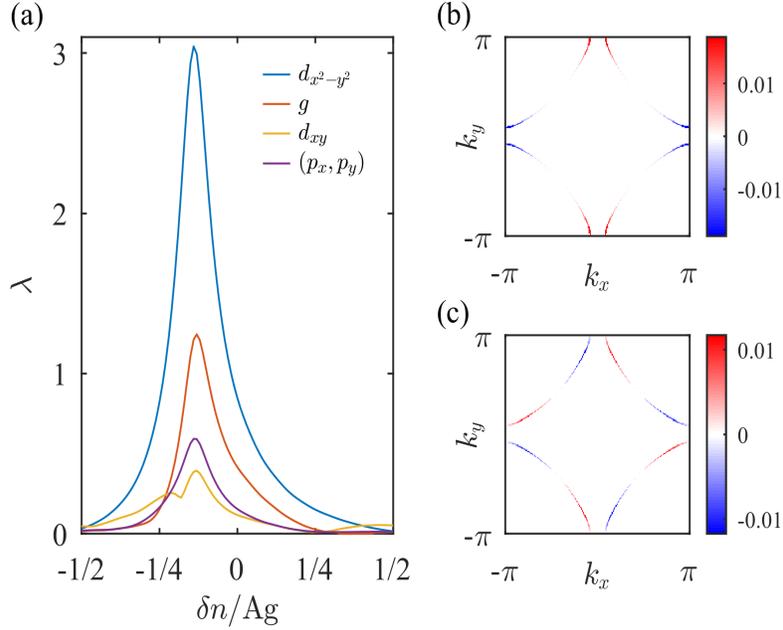

Figure 3. (a) Doping dependence of largest eigenvalue $\lambda$ for several pairing symmetry of flat $AgF_2$. Here $g$-wave is $xy(x^2-y^2)$ type. The order parameter $\phi(\mathbf{k})$ at Fermi surface of flat $AgF_2$ at an optimal doping $\delta n/Ag = -0.14$ for (b) leading pairing symmetry $d_{x^2-y^2}$ and (c) second leading pairing symmetry g-wave.

Under the fluctuation exchange approximation (FLEX), an effective pairing $\Gamma(q)$ between the electrons on the Fermi surface arising from spin and/or charge fluctuations can result in the formation of cooper pairs. The largest eigenvalue $\lambda$ of linearized Eliashberg equation become unity at superconducting $T_c$ and can imply the relative pairing strength near the $T_c$. Fig. 3(a) shows the doping dependence of largest eigenvalue $\lambda$ for several pairing symmetry of flat $AgF_2$ at T = 30 meV. These pairing symmetries are classified according to the character table of $D_{4h}$ group, symmetry group of the flat $AgF_2$. Here $g$-wave denotes $xy(x^2-y^2)$ type ($A_{2g}$ irrep). One can find that $d$-wave pairings for singlet channel have higher strength than $p$-wave pairings for triplet channel and the leading pairing symmetry is singlet $d_{x^2-y^2}$-type state ($B_{1g}$ irrep). Moreover, the electron doping readily decreases $\lambda$ while hole doping first tends to increase $\lambda$ reaching a peak value at an optimal doping of 14% and then decreases with further increase of hole doping concentration, consistent with the doping dependence of spin fluctuation (Fig. 2(b)). Fig. 3(b) and (c) plots show the order parameter $\phi(\mathbf{k})$ on the Fermi surface at an optimal doping of 14% for the leading pairing symmetry $d_{x^2-y^2}$ and sub-leading pairing symmetry $g$-wave. These findings are similar to the previous studies on cuprates, again highlighting the striking resemblance between $AgF_2$ and cuprates. [32,33] The present weak coupling approach is particularly suited to determine the main superconducting instability starting from the electron or hole overdoped regime. Near half-filling strong coupling effects become dominant and the superconducting phase is replaced by the magnetic insulating ordered phase which is known from previous experimental studies (cf. Ref. 3 and N. Bachar, et al. arXive:2105.08862). As for cuprates, one expects that the weak-coupling superconducting dome splits into two domes due to the suppression of superconductivity as the doping approaches half-filing. Thus, beyond weak coupling there should be an optimum doping level for hole and for electron doping as well. In the next two parts of this work, we suggest several chemical ways to execute electron or hole doping to the flat [$AgF_2$] sheet.

ELECTRON DOPING

In our previous study, RbMgF$_3$ emerged as the most promising substrate for epitaxial deposition of AgF$_2$, due to an optimal unit cell vector: large enough to prevent buckling of AgF$_2$ monolayer and small enough to ensure substantial enhancement of AFM interactions.[3] The structure of AgF$_2$-RbMgF$_3$ surface system is presented in fig. 4A. In this study, the RbMgF$_3$ substrate (labelled in short as RMF) consists of layers of MgF$_2$ stacked between layers of RbF. In order to dope the system in a controllable way, we substituted one of RbF with (MgO)$_2$ (in rocksalt structure) at different distances (depth) from AgF$_2$ monolayer. Fig. 4B shows an example of such system. We define $m$ as the number of stoichiometric RbMgF$_3$ layers separating the donor layer (MgO) from the acceptor layer (AgF$_2$). Since $m$ determines the distance between the layers we will use $m$ as our control variable hence in this example $m$ = 2. It should be emphasized that introduction of (MgO)$_2$ layer in between two adjacent MgF$_2$ layers in the structure is chemically reasonable and should be technically feasible in epitaxial setup due to similar chemistry and cationic-anionic separation in these compounds. Note on structural models: due to charge transfer occurring between reducing and oxidizing layers of the system (e.g. between MgO and AgF$_2$), a dipole moment emerges between those layers. Therefore, "sandwich" slabs equipped with a plane of symmetry in their middle were used in order to avoid formation of an artificial electric field necessary to fulfill boundary conditions in the direction parallel to the charge transfer and the slab ($z$). Consequently, Fig. 4 shows halves of these slabs in the middle panels, for clarity. The two example structures shown here are also provided in ESI as VASP output files.

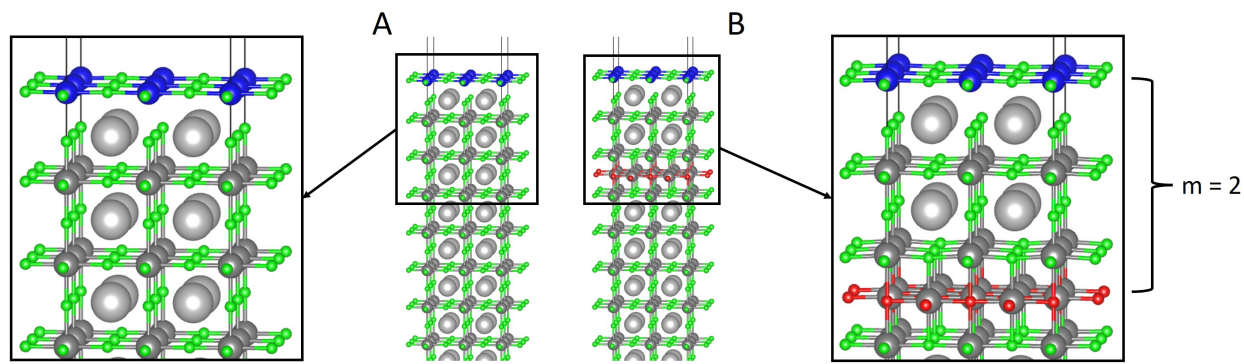

Figure 4. A) AgF$_2$ monolayer on RMF substrate (reference, undoped system), B) AgF$_2$ monolayer on RMF-MO substrate, with separation between AgF$_2$ and MgO $m$ = 2 (see text). Color code: Ag – blue, F – green, O – red, Rb – light grey, Mg – dark grey.

In order to assess the level of doping, we integrated and normalized electronic density of states (eDOS) in the upper Hubbard band (UHB) of AgF$_2$ monolayer on different RbMgF$_3$-MgO (RMF-MO) substrates, and took the fraction of occupied states in the UHB as the measure of doping level. As in the previous section, we label this value as $\delta n_{Ag}$ (positive for e-doping). The UHB in the reference RMF system with no oxide substitution is entirely empty, indicating $\delta n_{Ag}$ = 0 (fig. 5). As expected, for all doped cases the UHB is partially occupied. E.g., in the $m$ = 2 system, the UHB is occupied with $\delta n_{Ag}$ = 0.14 electrons per Ag atom. Thus, nominally neutral MgO is capable of serving as a charge reservoir $vs.$ AgF$_2$ due to the substantial difference of chemical potentials between them prior to charge transfer.[13]

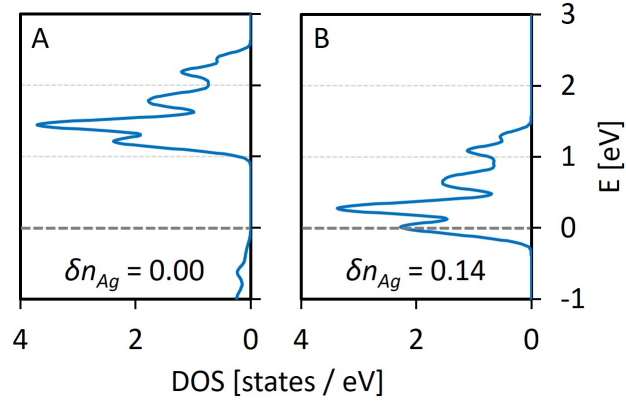

Figure 5. Comparison of eDOS plots of upper Hubbard band (UHB) of $AgF_2$ on pure RMF substrate (A) and on RMF-MO with *m* = 2 (B). Note the partial occupancy of the UHB in the latter system. "$\delta n_{Ag}$" refers to the level of electron doping, i.e. the fraction of occupied states in the UHB. Fermi level is marked with dashed gray line.

Fig. 6 shows the dependence of doping level on the separation between $AgF_2$ and MgO layers. As expected, the largest value of $\delta n_{Ag}$, 0.31, is observed for *m* = 0, i.e. the system in which the two layers are in direct contact. A similar value of 0.34 was obtained for $AgF_2$ in direct contact with bulk MgO.[3] This comparison shows that a single MgO layer adjacent to $AgF_2$ sheets suffices for overdoping. However, upon increasing separation, $\delta n_{Ag}$ decreases down to 0.07 for *m* = 5. In a simple capacitor model the charge transfer behaves as $\delta n_{Ag} \propto \frac{1}{m_0+m}$, with $m_0$ a constant[13]. Therefore, at larger separation, a change of *m* has a smaller effect on $\delta n_{Ag}$, as expected. Given the Ag/F vs. Cu/O analogy,[2,33] and similarity of their UHBs, it is expected that the optimum electron doping level for [$AgF_2$] sheet corresponds to ca. 0.15 just like in the case of the [$CuO_2$] one. Thus, our calculations suggest that the doping level to the former may be manipulated all the way from the underdoped regime (*m* = 4 or 3), via close to the optimum doping (*m* = 2), and up to the overdoped one (*m* = 1). In order to even more finely tune the level of doping and consequently other properties, we also investigated the case of RMF substrate in which **two** RbF layers have been substituted with $(MgO)_2$: one at *m* = 2 and additional one or two layers at *m* in the range from 3 to 5. For the resulting systems $\delta n_{Ag}$ falls in the 0.144-0.151 range as compared to 0.140 for *m* = 2. An analogous scan was also performed around *m* = 3. One may imagine that similar computational experiments may be performed for other geometrical arrangements of two, three and up to seven $(MgO)_2$ layers with various mutual placement of those, resulting in a possibility to dope the system of interest quasi-continuously in terms of $\delta n_{Ag}$ and in a rather broad range as requested for $T_C$ manipulations, just like in a capacitor/transistor setup with variable voltage. Importantly, atomic layer deposition techniques are capable these days to deposit such structures layer-by-layer.

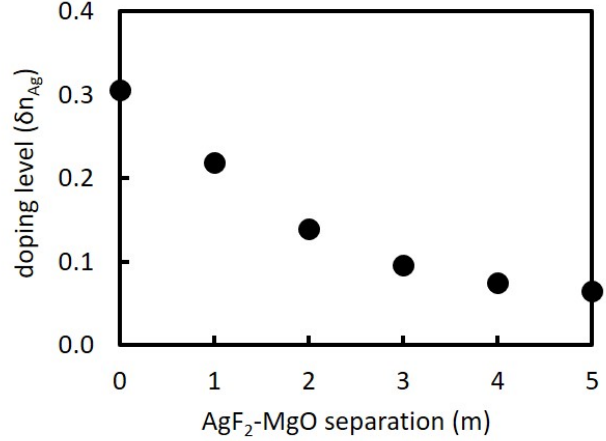

Figure 6. Dependence of e-doping level ($\delta n_{Ag}$) on number of separation layers between AgF$_2$ and MgO ($m$) in systems with a single MgO layer.

A number of structural changes in the AgF$_2$ monolayer are induced upon electron doping (fig. 7). Most noticeably, the distance between AgF$_2$ and the fluorine atom from the substrate increases with $\delta n_{Ag}$, which is expected, since an increased electronic density on Ag$^{II}$ cations due to doping should result in a weaker attraction between Ag$^{II}$ and F anions from the substrate. Electron doping also has a noticeable effect on the Ag-F-Ag angle in the monolayer. The general trend is that the angle becomes less straight with increasing $m$; however, upon doping of ca. 0.10, the monolayer becomes even flatter than in the undoped system. Ag-F bond length within the monolayer remains relatively stable up until ca. $\delta n_{Ag}$ = 0.22, whereupon the increasing puckering also leads to elongation of Ag-F bonds (Ag-Ag distance is constant and enforced by the unit cell size, which is the same for all values of $m$). The bond elongation, and concomitantly the decrease of the Ag-F-Ag angle, are naturally expected since electron doping corresponds to increased population of Ag-F σ* states.

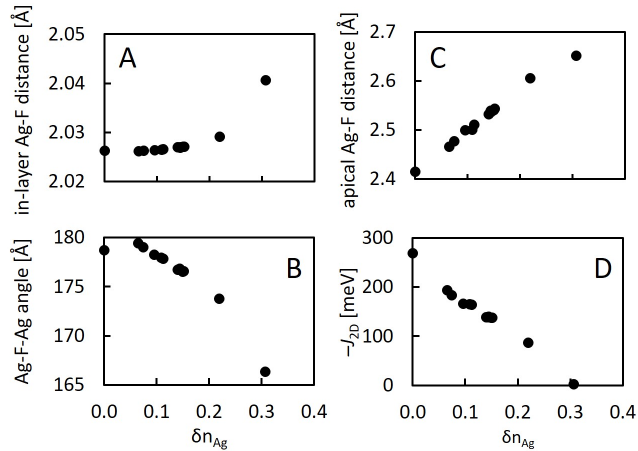

Figure 7. Dependence of selected parameters of AgF$_2$ monolayer on doping level: A) in-layer Ag-F distance, B) apical Ag-F distance (note, however, that for the largest level of doping here [$m$ = 0, $\delta n_{Ag}$ = 0.31], the atom in apical position is O and not F), C) Ag-F-Ag angle, D) AFM coupling constant ($-J_{2D}$).

Electron doping to UHB of AgF$_2$ has a strong influence on magnetic properties of the [AgF$_2$] sheet. For undoped, reference RMF-AgF$_2$ system, $J_{2D}$ is equal to ca. −270 meV (in agreement with previously published data).[3] This value decreases with increasing *x* in an approximately linear manner (fig. 7), eventually falling down to close to null (*ca.* 3 meV) for $\delta n_{Ag}$ = 0.31 (corresponding to *m* = 0). This matches well our previous results for AgF$_2$ on MgO substrate, which was even weakly ferromagnetic,[3] and with previous work on AgF$_2$-oxide superlattices[34,35] and overdoped cuprates.[36] Progressive decrease of the absolute value of $J_{2D}$ is driven by the tendency of free carriers to decrease the energy of the ferromagnetic state. An additional contribution originates in the increase of the Ag-F distance and departure of the Ag-F-Ag angle from 180° as expected from Goodenough-Kanamori-Anderson rules. Overall this decrease indicates that doping will destroy the long-range antiferromagnetic order but DFT is not accurate to determine the critical doping.

The attempts to more finely tune the level of doping and consequently other properties via introduction of two or more (MgO)$_2$ layers (as discussed above) allows also for fine tuning of the structural and magnetic properties. In the *m* = 2 example, $J_{2D}$ varies from −138 to −140 meV (as compared to −139 meV for *m* = 1) and the structural parameters differ by up to 1% from those for the singly substituted *m* = 2 system. These slight variations can be seen as an increased density of data point in fig. 7 around $\delta n_{Ag}$ ca. 0.15 and 0.10.

Having in mind our recent results regarding lattice relaxation and polaron formation in bulk and monolayer AgF$_2$,[12] we investigated the propensity for CDW formation at finite doping. To this end, we performed additional geometry optimization of the monolayer in *m* = 2 RMF-MO system, but with a flipped spin on the central Ag atom and with its four surrounding F atoms moved out of their high-symmetry positions, both in order to check for polaron formation. Since the doping level in this system is close to ⅛ (ca. 0.14), we utilized a larger supercell, with 8-AgF$_2$ formula units in the monolayer. The resulting solution features two distinct Ag centers: one with the average of four Ag-F bond lengths of ca. 2.038 Å and the other − 2.016 Å (as compared to 2.027 Å in the uniform solution) (fig. 8). Spin flip is not conserved upon optimization, but the solution is ferrimagnetic, since Ag atoms with elongated F contacts exhibit magnetic moments of ca. 0.35 $\mu_B$, while those with contracted contacts − 0.44 $\mu_B$, yielding non-zero net magnetization.

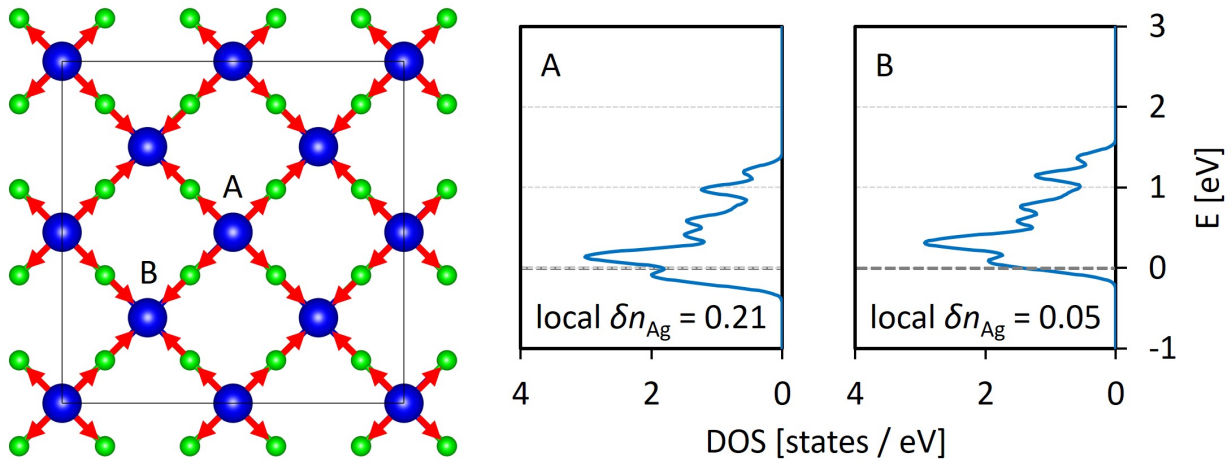

Figure 8. Left panel: AgF$_2$ monolayer in the disproportionated solution. Red arrows indicate displacement of F atoms relative to their position in uniform solution. Consequently, "A" and "B" indicate the two distinct Ag sites, with longer and shorter Ag-F contacts, respectively. Right panel: comparison of eDOS in UHB for the two Ag sites.

Electronic structure in fig. 8A & 8B indicates that some disproportionation does take place, with most of the additional charge (doping) localized in the UHB of one half of Ag sites (A; with longer Ag-F bonds), but the other half also features non-zero density of states at the Fermi level. Therefore, even though a distortion towards charge localization is marginally favorable, it is not strong enough to open the band gap and the charge is still distributed between all Ag sites, even if not in a perfectly uniform manner. This solution is more stable than the uniform solution only by ca. 7 meV/FU, and the divergence in Ag-F bond length between uniform and disproportionated solutions is only ca. 0.01 Å, which means that this distortion will likely be overcome by lattice vibrations at finite temperatures.

Despite its advantages,[2] RbMgF$_3$ can be regarded as quite an exotic substrate for atom layer deposition techniques, especially since it is currently unavailable commercially. Therefore, we have also investigated much simpler and readily available LiF and MgO (001) surfaces as substrates. Both compounds share the same structural type (rocksalt, $Fm\overline{3}m$), which could make such systems easier to obtain by epitaxial deposition than RbMgF$_3$-MgO systems, even despite their somewhat different unit cell sizes (LiF – 4.00 Å, MgO – 4.21 Å within our PBEsol models). Indeed, epitaxial deposition of LiF on MgO using pulsed-laser techniques has been reported in literature.[37]

Here, we have simulated two types of LiF-MgO heterostructures, which are presented in fig. 9. One of them uses LiF as bulk and is labelled LFMO (fig. 9A) and another – MgO (MOLF, fig. 9B). Both types of systems feature a varying number of LiF layers ($m$) separating AgF$_2$ monolayer and MgO. For LFMO, $m$ ranges from 1 to 3; for MOLF – from 1 to 4. In both cases, at least one layer of LiF separator is always needed to prevent overdoping of [AgF$_2$] sheets. To conserve computational resources, we investigated these systems in smaller cells without symmetry plane in $z$. We found out that due to the aforementioned electric field, which results from such an approach, the doping level $\delta n_{Ag}$ is consistently overestimated by the factor of ca. 1.1-1.2 with respect to symmetric "sandwich" cells. However, since we are concerned with the overall trend and the possibility of tuning of $\delta n_{Ag}$, rather than with its absolute values, we find this approximation to be acceptable.

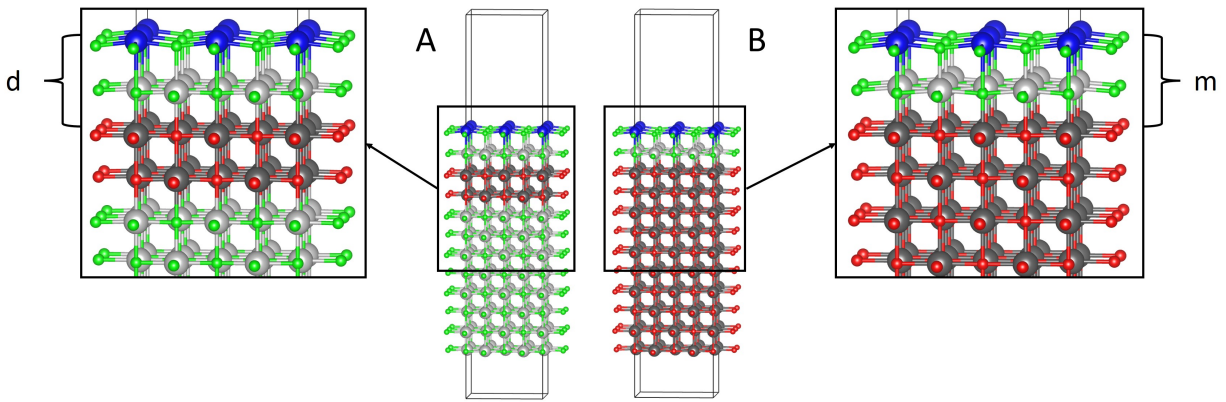

Figure 9. A) AgF$_2$ monolayer (flat solution) on substrate with LiF as bulk (LFMO). Here, two layers of MgO have been introduced to achieve doping. B) AgF$_2$ monolayer on substrate with MgO as bulk (MOLF). Here, one layer of LiF separator was used to prevent overdoping. Color code: Ag – blue, F – green, O – red, Li – light grey, Mg – dark grey.

The unit cell size of LiF is smaller than that of RbMgF$_3$, which, as previously reported, leads to destabilization of flat AgF$_2$ monolayer relative to puckered solution.[3] Also in the case of LFMO heterostructures, puckered solutions, with Ag-F-Ag angle of ca. 155°, are more stable than the flat solutions by 50-60 meV/FU (FU – formula unit) (similar to 156° and 60 meV/FU, respectively, on pure LiF in our previous work).[3] However, we have previously shown that the band structure and hopping parameters do not differ greatly between flat and puckered monolayers of $AgF_2$ and are still quite similar to those of cuprates,[3] more so than those of puckered layers in bulk $AgF_2$.[2] Therefore, study of $AgF_2$ on LFMO, even in puckered arrangement, is still worthwhile. It should be noted that in both LFMO and MOLF systems, F atoms from $AgF_2$ monolayer are strongly attracted to the LiF surface, resulting in Ag-F-Ag angle being in the 167-172° range even for "flat" (i.e. tetragonal) solutions.

Fig. 10 shows the dependence of AFM coupling constant on $x$ for $AgF_2$ monolayer on LFMO and MOLF substrates. Magnetic interactions are the strongest in flat solution on LFMO, but are lower by ca. 30-40 meV in the puckered solution, due to decreased overlap between Ag $d(x^2-y^2)$ orbitals and F $p$ orbitals parallel to Ag-F bonds.[2] In MOLF systems, $J_{2D}$ is even lower due to Ag-Ag distance being ca. 5% larger than in LFMO. Nevertheless, quite large $J$ values up to 130 meV (thus comparable to those measured for the cuprates) are achievable even for ca. $δn_{Ag}$ = 0.1 e-doped systems.

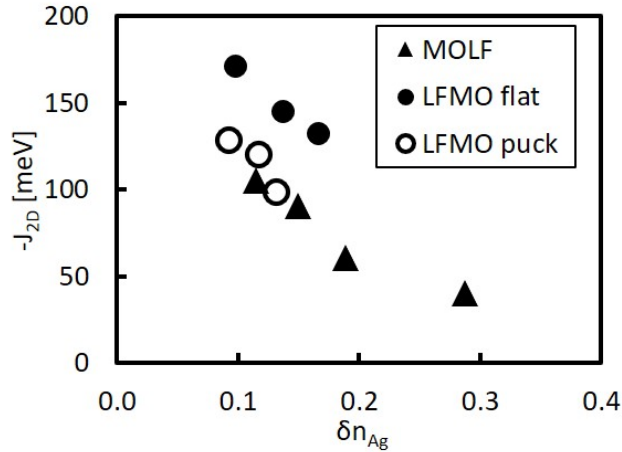

Figure 10. Dependence of AFM coupling constant ($-J_{2D}$) on doping level ($δn_{Ag}$) for MOLF and LFMO substrates. "puck" stands for puckered.

Overall, we find that the "chemical capacitor" approach[13] for e-doping of $AgF_2$, with appropriate substrates and experimental setup, is likely to provide access to a broad range of doping levels and thus $T_C$ values across the entire SC dome might be probed.

HOLE DOPING

In order to introduce holes into the monolayer of $AgF_2$ deposited on fluoride substrates, we employed three different strategies: (1) on-top fluorination of $AgF_2$ to a varying extent, (2) fluorination of a protective $Li_2F_2$ layer additionally deposited on top of $AgF_2$ monolayer, again to a varying extent, and (3) bonding with a strong oxidizer: $PtF_2$ or $KrF_2$ in the "chemical capacitor" fashion. We show that in approach (1) there is a tendency toward depopulation of Ag $d(z^2)$ orbital, with formation of $Ag^{III}$ high-spin ion, while the bandgap retains a non-zero value. In (2) and (3), the resulting eDOS show electronic states appearing at the Fermi level, again originating predominantly from Ag $d(z^2)$ orbitals. For h-doped systems obtained here, it has proven difficult to evaluate doping level ($δn_{Ag}$) from eDOS, due to lack of separation of bands of interest from the remaining electronic states. Therefore, we use instead the stoichiometric degree of fluorination or oxidation, i.e. the fraction of Ag atoms connected to F or strong oxidizer molecule, designated as *x*.

(1) Direct fluorination of AgF$_2$ monolayer

Direct fluorination of flat AgF$_2$ monolayer was performed at three different levels: ¼, ½ (with two different geometric combinations) and 1, corresponding to AgF$_{2.25}$, AgF$_{2.5}$ and AgF$_3$, respectively. Similarly, as for e-doping, these scenarios were realized on two different substrates: RbMgF$_3$ (RMF) and LiF (LF). As mentioned before, while RMF has been previously found to have an optimal unit cell size and is thus useful as model system,[3] it is not commercially available. Therefore, in order to explore experimentally-possible path of hole-doping via fluorination, we also conducted fluorination of initially puckered (LF$^{(I)}$) and flat (LF$^{(II)}$) AgF$_2$ monolayer on readily available LiF substrate; bearing in mind that the puckering of AgF$_2$ monolayer is energy-preferred on this substrate.[3] All calculations were carried out with the emphasis to find the energy-preferred spin arrangement between metal centers (AFM or FM), and the possibility of occurrence of Ag$^{III}$ low-spin state (non-magnetic), when fluorine atoms are attached.

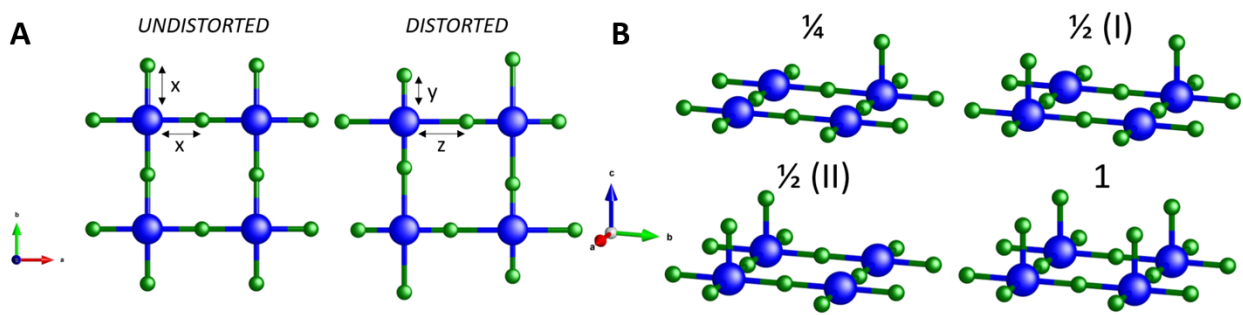

Figure 11. A – undistorted and distorted monolayer on RMF – starting point for optimization, B – fluorination patterns of AgF$_2$ monolayer. Color code: Ag - grey, F-green.

Fig. 11B shows patterns of direct fluorination. ¼ fluorination refers to apically bonding of F atom to every fourth silver atom. For the ½ case, where every second Ag atoms is fluorinated, two combinations were explored – diagonal (I) and along the *a* vector (II) in the monolayer. Complete fluorination ($x = 1$) effectively leads to the formation of a monolayer with AgF$_3$ stoichiometry. For both substrates (RMF and LF) and for all stoichiometries and patterns, direct fluorination reaction is favorable in terms of *ΔG* (cf. ESI).

In addition, in order to check the propensity towards distortion of fluorinated AgF$_2$ monolayer, we also optimized initially distorted and undistorted AgF$_x$ structures separately on the RMF substrate. Fig. 11A presents distorted and undistorted sheets of AgF$_2$. The **x**, **y** and **z** values were dependent on the unit size of substrate; and for optimized AgF$_2$ monolayer placed on RMF and LF were equal to 2.03 Å, 2.00 Å (LF$^{(1)}$) and 2.02 Å (LF$^{(2)}$), respectively. Distortion was implemented as alternate elongation and contraction of Ag-F in-layer bonds (denoted as **y**, **z**) by 0.30 Å from their initial value (**x**), observed in the undistorted structure. However, all of these distorted structures either converged to uniform solution upon relaxation, or were found to be higher in energy than the uniform solution.

(2) Indirect fluorination via a protective LiF monolayer

Alternatively, to avoid direct fluorination of AgF$_2$, we investigated fluorination of a protective intermediary LiF layer, placed on top of the AgF$_2$ monolayer (in either a puckered or flat arrangement of the latter). To ensure sufficiently short distance (and resulting charge transfer) between AgF$_2$ and the fluorinated LiF layer (hole source), the choice of LiF as a substrate was reasonable due to its shorter unit cell size. We investigated fluorination of Li$_2$F$_2$ layer (with AgF$_2$ as a formula unit) with $x$ = ¼, ½ and 1, as well as a reference system without fluorination (fig. 12). We investigated the system in both flat and puckered arrangement of AgF$_2$; only puckered AgF$_2$ is shown in fig. 12. Unfortunately, in contrast to the first strategy, fluorination of protective LiF layer on top of AgF$_2$ is uphill in free energy for all investigated stoichiometries.[38] This should not come as a surprise: LiF is an ionic solid, and its constituent Li(I) cations, with their remaining two 1s core electrons, have very little propensity towards such reaction. However, properties of those hypothetical systems are still worth looking into, and they will be discussed in further sections.

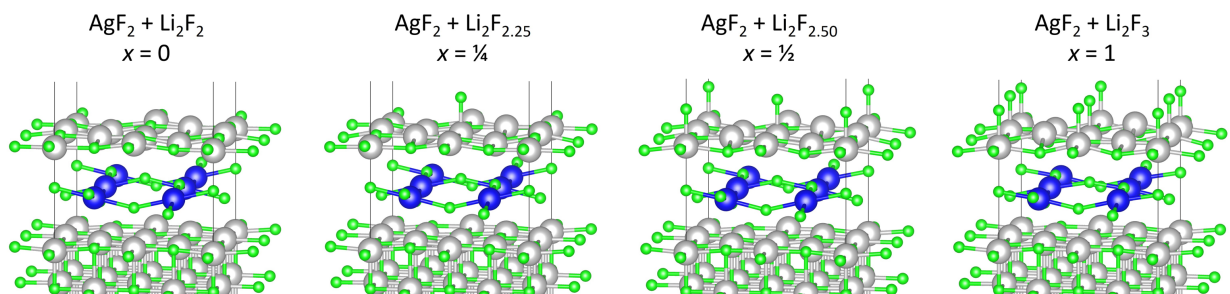

Figure 12. Optimized AgF$_2$+Li$_2$F$_{2+x}$ structures for antiferromagnetic ordering between metal centers. Color code: Ag – blue, F – green, Li – grey.

(3) Bonding with strong oxidizers

In search for a convenient and practical way of hole-doping to AgF$_2$ monolayer, we also considered two molecules of strong oxidizers: PtF$_6$ and KrF$_2$, as electron sinks for doping. Again, RbMgF$_3$ (RMF) served as a substrate for the AgF$_2$ monolayer.[3] The view of structures was presented in the fig. 13. KrF$_2$ was placed on every second (A) or each (B) silver atom. PtF$_6$, because of its larger volume, was placed only on every second silver atom (C), in order to avoid steric hindrance. Adsorption of KrF$_2$ in both concentrations is thermodynamically favorable (negative ΔG), while the reaction with PtF$_6$ is slightly uphill.

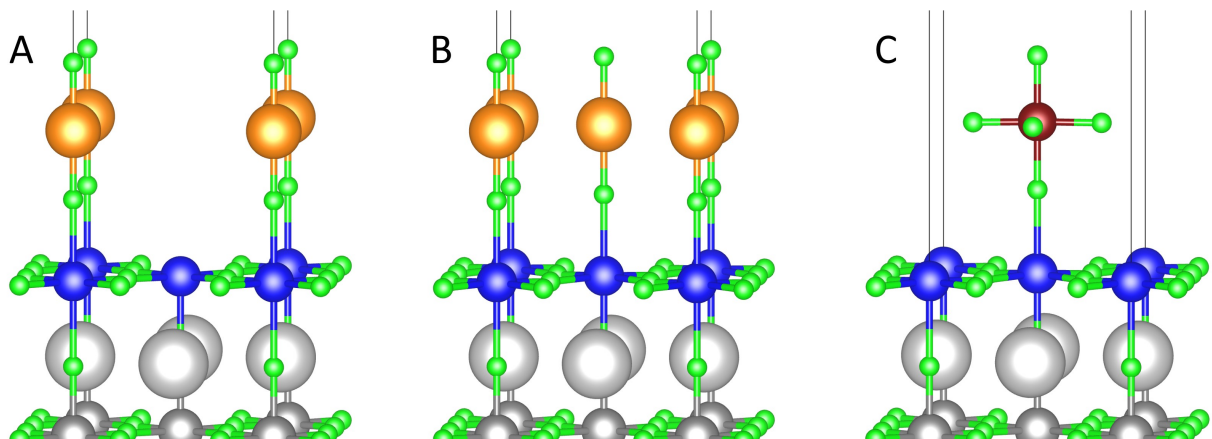

Figure 13. Strong oxidizer molecules bonded to AgF$_2$ monolayer. A and B – KrF$_2$ with $x$ = ½ (A) and 1 (B), C – PtF$_6$ with $x$ = ½. Color code: F – green, Mg – dark grey, Ag – blue, Rb – light grey, Kr – orange, Pt – brown. Rb-F bonds are not shown for clarity.

*Structural properties*

As was the case in e-doping, introducing holes using the above strategies leads to structural changes to the $AgF_2$ monolayer, and we can identify some common trends among them. In general, we should note that Ag-F-Ag angle in the monolayer is a very good reduced parameter for following structural changes: since the Ag…Ag distance in the monolayer is fixed and determined by a given substrate, Ag-F-Ag angle is also directly related to the Ag-F bond length.

For flat monolayer on an appropriately-sized substrate (RMF), both fluorination as well as reaction with a strong oxidizer generally lead to repulsion of in-layer F atoms away from the fluorine/oxidizer and towards the substrate. This can be seen as divergence of Ag-F-Ag angle from close to 180 degrees to lower values (fig. 14). The dependence is monotonic for directly fluorinated $AgF_2$, while for oxidizers, the effect is present in reaction with $PtF_2$ and for $KrF_2$ with $x = 1$. We can interpret this distortion as a result of repulsive interaction between F p orbitals of $AgF_2$ which are perpendicular to the monolayer, and dopant F p orbitals orthogonal to the axial bonds between Ag and fluorine/oxidizer molecule.

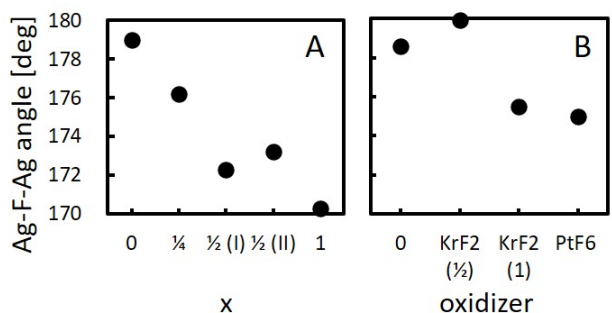

Figure 14. Dependence of Ag-F-Ag angle (average value) on: A – fluorination level for the directly fluorinated $AgF_2$ monolayer, B – type of oxidizer molecule bonded to the $AgF_2$ monolayer.

The same effect, i.e. repulsion of in-layer F atoms towards the substrate, is seen in the fluorinated LF system. However, since the $AgF_2$ monolayer is puckered in its ground state, this change does not result in a dependence like in fig. 14. Rather, it is manifested as a change in fraction of F atoms above and below the plane of Ag atoms. (fig. 15). Consequently, for $x = ½(I)$ and $x = 1$ levels of direct fluorination, initially puckered and flat monolayers converge to the same solution. (In the flat $AgF_2$ solution on LF, all F atoms in the monolayer are also shifted towards the substrate relative to Ag atoms, which is a way of releasing strain the monolayer experiences while atop of LiF.[3]) This can also be seen in the dependence of relative stability of flat and puckered $AgF_2$ monolayer on LF (fig. 15). Flat arrangement becomes less unstable with increasing fluorination level, or rather, the two solutions become more similar to each other, and therefore closer in energy.

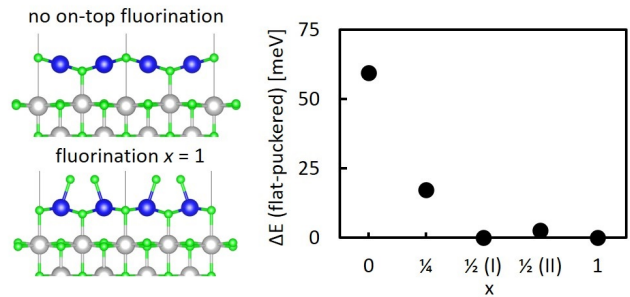

Figure 15. Left panel – comparison of puckered arrangements of AgF$_2$ monolayer on LiF with and without fluorination. Right panel – dependence of relative stability of flat and puckered monolayer in LiF system on fluorination level *x*.

A slightly different trend is seen in indirectly fluorinated systems on LF (via protective LiF monolayers). Fluorination decreases the energy difference between puckered and flat AgF$_2$ solutions (fig. 16); however, the monolayer in the puckered solution does not experience the kind of distortion visible in case of direct fluorination. Actually, upon complete indirect fluorination (*x* = 1), the corrugated monolayer becomes somewhat flatter, as evidenced by the dependence of Ag-F-Ag angle on *x* (fig. 16). Conversely, it is the flat solution that becomes more corrugated, starting from almost ideally flat (179.5°) at *x* = 0 to deformed in a similar way as in the case of complete direct fluorination. As mentioned above, "naked" AgF$_2$ in flat arrangement (i.e. tetragonal symmetry) on LF is prone to a tetragonally symmetrical corrugation. Here, an additional protective LF monolayer on top enforces a flatter arrangement of AgF$_2$ in the flat solution, but it leads to an increased strain within it, resulting in larger ΔE between solutions, compared to "naked" AgF$_2$ monolayer on LF.

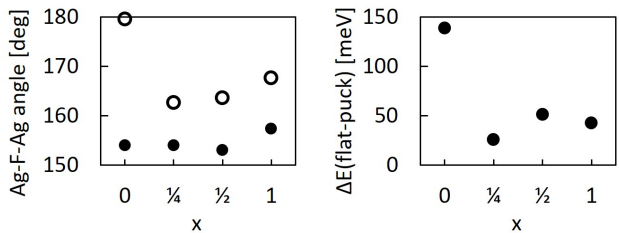

Figure 16. Left panel: dependence of Ag-F-Ag angle (average value) for puckered (solid) and flat (hollow) solutions. Right panel: relative stability of flat and puckered AgF$_2$ (right panel) in systems with indirect fluorination.

In the following section, where magnetic and electronic properties are discussed in systems on LF substrate, only the puckered monolayer of AgF$_2$ (i.e. the ground state) is considered.

*Electronic and magnetic properties*

AgF$_2$ monolayers are characterized by AFM ordering in the undoped state.[3] Fluorination or reaction with strong oxidizers brings changes to its magnetic properties. Direct fluorination leads to formation of distinct high-spin Ag$^{III}$ states on the sites where F atoms are attached, which is indicated by an increased magnetic moment, compared to Ag$^{II}$ (fig. 17); localization of extra holes as Ag$^{III}$ is also seen in a different coordination sphere of Ag$^{III}$ as compared to Ag$^{II}$. Moreover, some magnetic moment is also present on apical F atoms attached to the surface. With the increasing degree of fluorination, the local magnetic moment on Ag$^{III}$ and Ag$^{II}$ decreases. The same trend is found for LF system (not shown).

Conversely, for indirectly fluorinated AgF$_2$ on LF, magnetic moment on Ag sites remains uniform (ca. 0.52 µ$_B$) and does not vary with fluorination degree.

We expect that the magnetic structure to be very sensitive to the choice of parameters (Hubbard U and Hund's rule J). Therefore, the high-spin states found should be taken with a pinch of salt and checked with a more accurate quantum chemistry method.

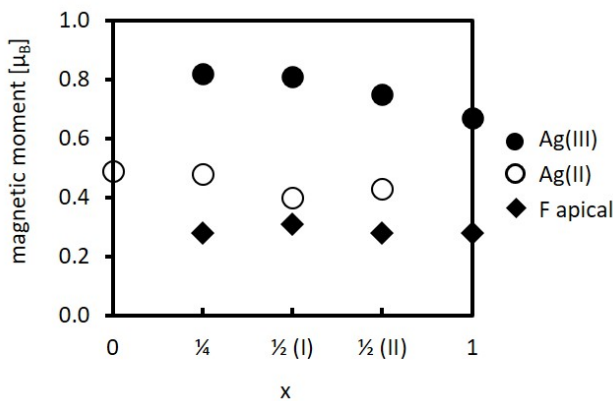

Figure 17. Dependence of magnetic moment on fluorination degree (*x*) in directly fluorinated RMF-AgF$_2$ system.

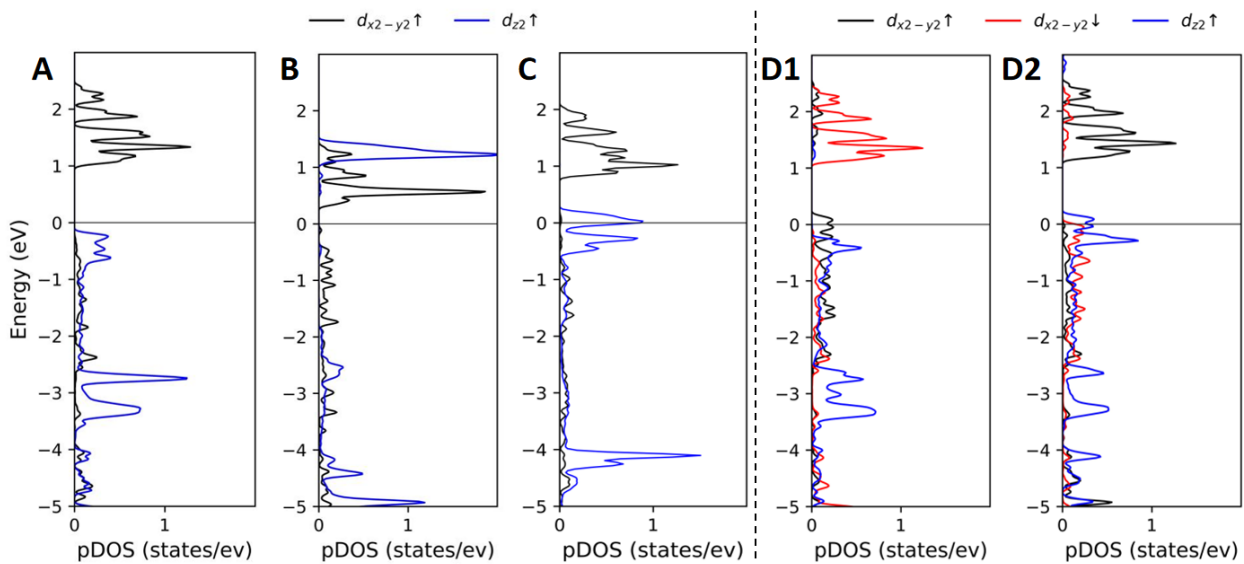

Figure 18. Electronic structure of Ag states in selected, representative hole-doped systems. A – RMF-AgF$_2$ (undoped, reference), B – RMF-AgF$_3$ (complete fluorination, *x* = 1), C – LiF-AgF$_2$-Li$_2$F$_3$ (complete indirect fluorination, *x* = 1), D – RMF-AgF$_2$-PtF$_6$, states from unbonded silver atoms (D1) and bonded silver atoms (D2) to PtF$_6$ molecule.

From the point of view of HTSC, the most interesting property of fluorinated and oxidized systems considered here is their electronic structure. A comparison of eDOS in systems resulting from the three strategies employed in this work is shown in fig. 18.

For the directly fluorinated AgF$_2$, the band gap retains a non-zero value for all stoichiometries (example shown in fig. 18B). RMF exhibits a narrower bandgap (0.96eV) than LF (1.27 eV), due to flat arrangement of the monolayer and consequently stronger orbital overlap.[3] Nevertheless, for both substrates, the band gap decreases with fluorination degree (cf. ESI). In the final step (*x* = 1), bandgaps are as narrow as 0.12 and 0.24 eV for RMF and LF, respectively. The character of the valence band undergoes modification upon fluorination. As additional F atoms are apically bonded to Ag$^{II}$, at the vicinity to Fermi level, new states arise. The new bands, just below Fermi level, are narrow and originate from Ag$^{III}$ and F (apical) atoms, with a majority contribution of the latter (cf. ESI). Taken together with non-zero magnetic moments on additional F atoms, it appears that the chemical species introduced upon fluorination has a mixed Ag$^{III}$ and F$^{\bullet}$ radical character. Most importantly, the holes are not distributed evenly between Ag sites, but localized on sites of additional F bonding. Hence, this approach unfortunately does not lead to proper doping of AgF$_2$ flat layers.

The situation is different for indirect fluorination (fig. 18C). When additional fluorine atoms are attached ($x = \frac{1}{4}$) to Li$_2$F$_2$, Fermi level lowers. This leads to depopulation of Ag states, to an extent which increases with fluorination level. The greatest effect is noticeable for $x = 1$, with even a slight deformation of UHB (between 0.9-2.0 eV). More insightful orbital-resolved analysis, reveals that states passing Fermi level constitute mainly Ag d(z$^2$) orbitals. These states are strongly admixed with the p(z) states from apically bonded fluorine atoms (from on-top LiF layer) (cf. ESI). Approach of Li$_2$F$_2$ atop layer fluorination ensures the non-zero value of eDOS at the Fermi level, which together with uniform values of magnetic moments at Ag sites indicates a delocalized distribution of holes. However, these new empty states near the Fermi level are a result of depopulation of Ag d(z$^2$) orbitals, which are perpendicular to the AgF$_2$ monolayer and not involved in the intra-sheet AFM interactions. Thus, regretfully, also this approach does not correspond to a desired proper doping.

Finally, for the systems with AgF$_2$ monolayer bonded to strong oxidizers, states passing through the Fermi level also arise. However, this is the case only in the system with adsorbed PtF$_6$. Only half of Ag sites are bonded to PtF$_6$, but both sites exhibit almost the same distribution, within the energy range from −3 to 3 eV. The difference of local magnetic moment between these atoms reaches ca. 0.13 $\mu_B$. Orbital-resolved analysis for d(x$^2$-y$^2$) and d(z$^2$) reveals major contribution of the latter to total DOS, right above the Fermi level.

The band structure close to the Fermi level is simple enough to allow for a direct computation of the number of holes in the AgF$_2$ monolayer using a tight-binding fit and Luttinger sum rule. Formally Pt has a 5d$^4$ configuration and it is fully polarized. One of the holes in the 5d-shell is actually shared in three fully polarized bands. Two bands are associated to the d(yz) and d(xz) orbitals of the Pt atoms hybridized with fluorine p orbitals and hence represent charge close to the dopant. The remaining band is formed by mixed polarized Ag(d)-F(p) in-plane orbitals. It host $\delta n_{Ag}$ = −0.10 in-plane hole doping. However, it is likely that the strong ferromagnetic character of the Pt will frustrate superconductivity or, at most, favor a p-wave order parameter. In any case, this approach also fails to deliver a proper h-doped AgF$_2$ layer analogous to cuprates.

CONCLUSIONS

Our results indicate that the analogy between a single layer of AgF$_2$ and flat CuO$_2$ sheets found in parent compounds of oxocuprate superconductors extends even to the type and level of doping which result in maximum $T_C$ for both systems, and to the pairing symmetry. Although numerous similarities between both families of compounds have been postulated early on,[39] those detected in the current work render a single AgF$_2$ layer probably the best analogue of copper oxides found so far, and despite many previous attempts.[40,41] Following these findings, we show that electron doping to AgF$_2$ can be realized in systems where fluoride-oxide heterostructures are used as substrates for deposition of $AgF_2$. Such systems fulfill two purposes simultaneously: a) stabilization of monolayer in a nearly flat arrangement to maximize 2D antiferromagnetic interactions, as previously reported,[3] and b) controlled injection of electrons into the monolayer to provide a desired level of electron-doping, as reported here. In all of the studied systems, doping level in the range 0.1-0.3 appears to be easily accessible, raising hopes for finding a superconducting dome in the phase diagram. The number of the charge-reservoir oxide layers and their separation from the [$AgF_2$] sheet turn out to be key ingredients of the recipe for doped-$AgF_2$. LiF-MgO heterostructures appear to be the most promising for future experiments because of their simple structure, relatively similar chemistry, reasonable match with the lattice constant of tetragonal $AgF_2$ layer, and commercial availability of good quality optical crystals. Preparations for experimental verification of these findings are ongoing. Conversely, approaches for hole doping in $AgF_2$ monolayer which have been investigated here did not produce a desirable outcome, in the way that the resulting depopulation of Ag d states was localized predominantly on the $d(z^2)$ orbital, which is perpendicular to the layer and not involved in 2D AFM interactions. However, those h-doped systems in which a non-zero eDOS arises at the Fermi level are also characterized by unfavorable enthalpy of fluorination/oxidation. This is in line with previous findings regarding work function of $AgF_2$, which make hole doping in this compound a very challenging feat in general.[7] It remains worthwhile to search for other approaches, some of which could prove more promising in the future, since as we have shown, achieving $δn_{Ag}$ = −0.14 (h-doping) would be the most desirable outcome. However, superconducting oxocuprates are also found in the e-doping regime, and the chemical capacitor strategy for e-doping $AgF_2$ we have investigated here has a great potential for successful implementation.

ACKNOWLEDGMENTS

WG thanks the Polish National Science Center (NCN) for the Maestro project (2017/26/A/ST5/00570). This research was carried out with the support of the Interdisciplinary Centre for Mathematical and Computational Modelling (ICM), University of Warsaw under grant ADVANCE++ (no. GA76-19) and SAPPHIRE (no. GA83-34, G85-892). JL acknowledges financial support from Italian MIUR through Project No. PRIN 2017Z8TS5B, and from Regione Lazio (L. R. 13/08) through project SIMAP. M.N.G. is supported by the Marie Sklodowska-Curie individual fellowship Grant agreement SILVERPATH No: 893943. L.X.-Q., S.K.P. and J.F. thank the National Natural Science Foundation of China (Grant No. 11725415 and No. 11934001), the Ministry of Science and Technology of China (Grant No. 2018YFA0305601 and No. 2016YFA0301004), and the Strategic Priority Research Program of the Chinese Academy of Sciences (Grant No. XDB28000000).